\magnification=1200
\baselineskip=18truept
\input epsf

\def\preprint{Y}
\def\draftversion{N}
\def\cap{\hsize=4.6in}

\if \draftversion Y


\fi
\def\figure#1#2#3{\vskip .2in
\if \preprint Y \midinsert \epsfxsize=#3truein
\centerline{\epsffile{figure_#1_eps}} \halign{##\hfill\quad
&\vtop{\parindent=0pt \hsize=5.5in \strut## \strut}\cr {\bf Figure
#1}&#2 \cr} \endinsert \fi}

\def\figureb#1#2{\if \preprint N \midinsert \epsfxsize=#2truein
\centerline{\epsffile{figure_#1_eps}} \halign{##\hfill\quad
&\vtop{\parindent=0pt \hsize=5.5in \strut## \strut}\cr \cr \cr
\cr \cr \cr  {\bf Figure #1} \cr} \endinsert \fi}
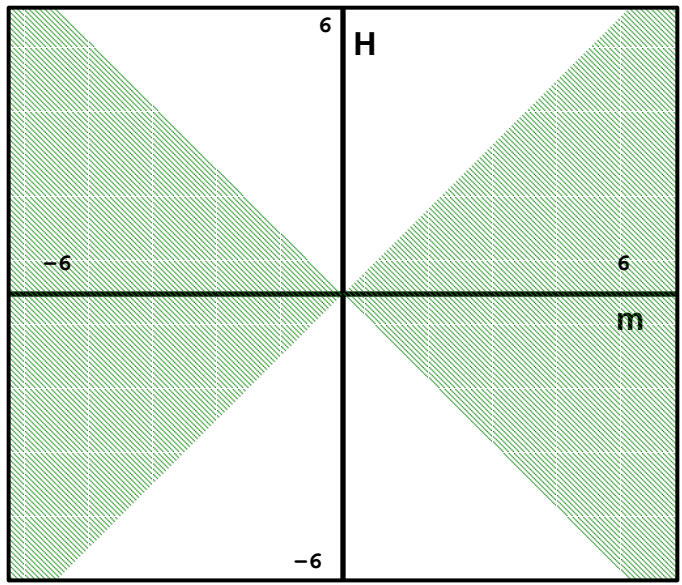
\def\captionone{\cap Spectrum of the Dirac Hamiltonian in the continuum.
All oblique lines have slopes $\pm 1$.}
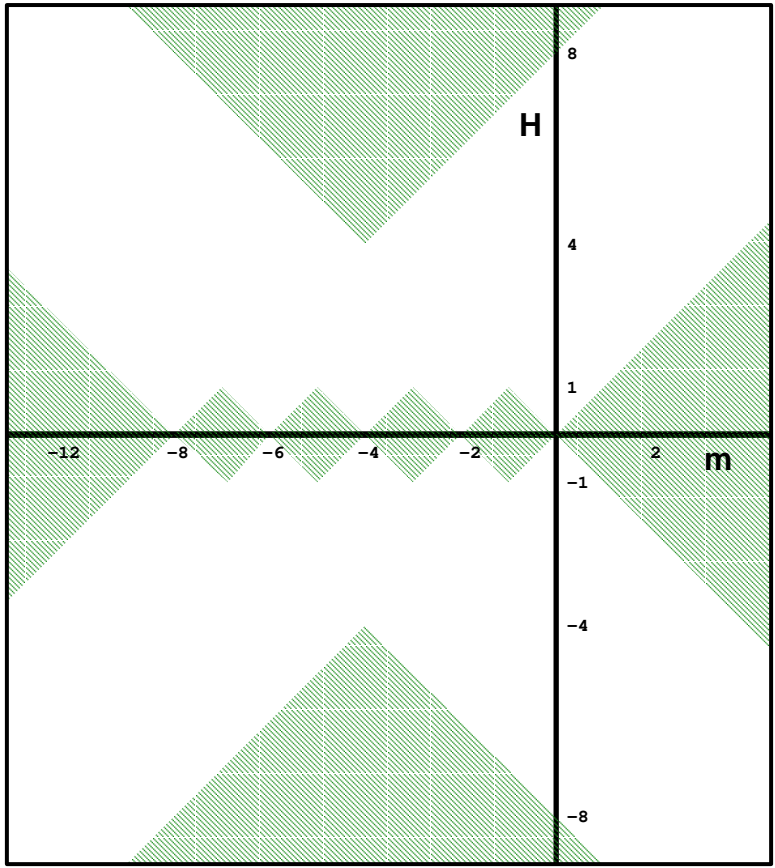
\def\captiontwo{\cap Spectrum of the Wilson Dirac Hamiltonian on the lattice 
for d=4. All oblique lines have slopes $\pm 1$.}
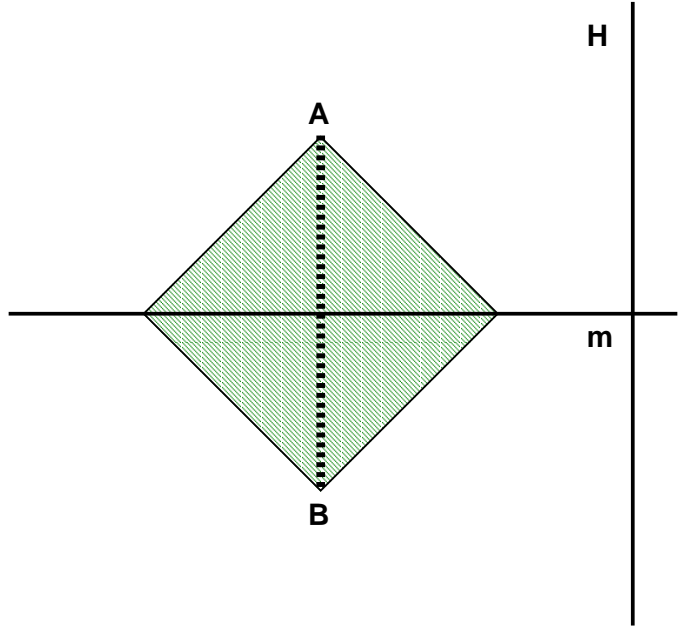
\def\captionthree{\cap A rhombus containing no eigenvalues. 
All oblique lines have slopes $\pm 1$.}
\def\Journal#1#2#3#4{{#1} {\bf #2}, #3 (#4)}


\def\NPB{{\it  Nucl. Phys.} B}
\def\PLB{{\it  Phys. Lett.}  B}
\def\PRL{\it  Phys. Rev. Lett.}
\def\PRD{{\it  Phys. Rev.} D}

\def\LTP{{\it Nucl. Phys.} B {(Proc. Suppl.)}}
\def\SCRI{1}
\def\horn{2}
\def\vanbaal{3}
\def\reedsimon{4}
\def\ovlap{5}
\def\wupp{6}
\def\kerler{7}
\def\hernan{8}
\def\adams{9}
\def\qek{10}
\def\ek{11}
\def\schwinger{12}
\def\boulder{13}
\def\urspp{14}
\def\urscheck{15}
\def\prddwf{16}
\def\columa{17}
\def\columb{19}
\def\stama{18}
\def\morning{20}
\def\ape{21}
\def\steph{22}
\def\chiraltd{23}

\line{\hfill RUHN-99--4}
\vskip .5cm
\centerline {\bf Bounds on the Wilson Dirac Operator.}
\vskip 1cm
\centerline{Herbert Neuberger}
\vskip .25cm
\centerline{\tt neuberg@physics.rutgers.edu}
\vskip 1.5cm
\centerline{\it Department of Physics and Astronomy}
\centerline{\it Rutgers University}
\centerline{\it Piscataway, NJ 08855--0849}
\vskip 2cm
\centerline{\bf Abstract}
\vskip 2cm
New exact upper and lower bounds are derived on the spectrum of the square
of the hermitian Wilson Dirac operator. It is hoped that the derivations
and the results will be of 
help in the search for ways to reduce the cost of simulations 
using the overlap Dirac operator. The bounds also apply to the
Wilson Dirac operator in odd dimensions and are therefore relevant
to domain wall fermions as well. 

\vfill\eject
{\bf Introduction}
\vskip .5cm
Let $D (m)$ denote the continuum Euclidean Dirac operator where
the real parameter $m$ is the fermion mass. 
In even dimensions $d$ 
a generalization of $\gamma_5$ exists and 
shall be denoted by $\gamma_{d+1}$. $D (0)$ is antihermitian
and anticommutes with $\gamma_{d+1}$. 
Then, $H(m)=\gamma_{d+1} D(m)$ is
hermitian. $H(m)$ will be referred to as the hermitian Dirac operator. 
A characteristic property
of this operator is the range of its spectrum as a function of the real
mass parameter $m$. Since $H^2(m) = D^\dagger (m) D(m)$ it is meaningful
to consider the spectrum of $H^2(m)$ both in even and odd dimensions. 

Figure 1 displays the familiar spectral structure 
of $H(m)$ in the continuum in an arbitrary fixed gauge background. The 
boundaries 
shown come from rigorous lower bounds on the spectrum of $H^2 (m)$.
These bounds hold for any gauge background and are often saturated,
for example in the case that the gauge background is trivial, or in 
the case that
it consists of a gauge field carrying non-zero topology. There is no upper
bound on $H^2 (m)$, and the spectrum will indeed increase indefinitely
in any fixed smooth gauge background. All this holds also 
on a compact manifold, henceforth taken to be a flat torus.

\figure{1}{\captionone}{3.5}

The objective of this paper is to clarify what happens when the massive
Dirac Hamiltonian is put on the lattice following Wilson's prescription.
The most fundamental feature of a lattice operator is that its spectrum
is absolutely bounded from above - this is how the lattice acts as
a regulator. However, lower bounds obeyed by the hermitian 
Wilson Dirac operator, $H_W^2 (m)$, are also very important, because
often we wish to use $H_W (m)$ to put massless, or almost massless
quarks on the lattice.

When the gauge background is trivial, $H_W (m)$ can
be explicitly diagonalized and one finds the spectral structure shown
in Figure 2. A simpler derivation is contained in what follows.

\figure{2}{\captiontwo} {4}

When the gauge field is turned on the figure gets distorted. The
upper bound on $H_W^2 (m)$ remains unchanged, and so does the lower
bound for positive values of the mass parameter $m$. Changes
occur only for $m < 0$ and for the lower bound of $H_W^2 (m)$.
So long
we are close to the trivial case the distortion is small: it amounts
to the replacement of the string of rhombi in Figure 2 by a string of
smaller rhombi, inscribed into the ones we have in Figure 2. The new
rhombi no longer touch each other. As $m$ is varied, 
eigenvalues of $H_W (m)$ can cross zero in the 
intervals that open up, separating the rhombi. When the gauge background is 
random enough the internal rhombi close up completely and very
low eigenvalues of $H_W^2(m)$ are no longer excluded [\SCRI] 
for any mass in the segment $(-2d,0)$. 
For any
gauge background the figure
stays mirror symmetric about the $m=-d$ vertical line. 

Although we focus on even dimensions here, so long we phrase the results
for the Wilson Dirac operator itself and not its hermitian version,
they hold for odd dimensions as well. In particular, the five dimensional
case applies to domain wall formulations of QCD. 

\vskip .5cm

{\bf Notations and Conventions}
\vskip .5cm
Let us start by establishing our notation. 
We are working on a $d$-dimensional
hypercubic lattice. When comparing to the continuum
the lattice spacing is denoted by $a$. 
On its links we have $SU(n)$ matrices
$U_\mu (x)$ which make up the gauge background the fermions
interact with. $\mu = 1,2,\dots d$ denotes positive directions
and $x$ denotes a lattice site. The lattice is finite. 

The fermions are vectors $\psi_\alpha^i (x)$. $\alpha$ is
a spinorial index, $i$ is a gauge group index and $x$ is a
lattice site. The action on the fermions
is described in terms of several unitary operators.
First are the Euclidean Dirac $\gamma_\mu$'s which
act only on spinorial indices. Second come 
the directional parallel transporters $T_\mu$
which act on the site index and the group index. They are defined
by:
$$
T_\mu (\psi ) (x) = U_\mu (x) \psi (x+\hat\mu) .
$$
A third class of unitary operators implements gauge transformations,
each characterized by a collection of $g(x)\in SU(n)$
acting on $\psi$ pointwise, and only on the group
indices. The action is represented by a unitary operator 
$G(g)$ with $(G(g)\psi)(x) =g(x)\psi(x)$. The $T_\mu$ operators 
are ``gauge covariant'',
$$
G(g) T_\mu (U) G^\dagger (g) = T_\mu (U^g ) ,
$$
where,
$$
U^g_\mu (x) = g(x) U_\mu (x) g^\dagger (x+\hat\mu ) .
$$
The variables $U_\mu (x)$ are distributed according to a
probability density that is invariant under $U\to U^g$ for any $g$. 

The lattice replacement of the massive continuum Dirac operator, $D(m)$,
is an element in the algebra generated 
by $T_\mu,~ T_\mu^\dagger ,~ \gamma_\mu$.
Thus, $D(m)$ is gauge covariant. For $U_\mu (x) = 1$ the $T_\mu$ become
commuting shift operators. 

The Wilson Dirac operator,
$D_W (m)$ is the sparsest possible analogue of the continuum massive
Dirac operator which obeys hypercubic symmetry. 
Fixing the so called $r$-parameter
to its preferred value ($r=1$),
$D_W (m)$ can be written as:
$$
D_W = m+\sum_\mu(1- V_\mu );~~~~V_\mu^\dagger V_\mu =1;~~~V_\mu=
{{1-\gamma_\mu}\over 2} T_\mu +{{1+\gamma_\mu}\over 2} T_\mu^\dagger .
$$
In even $d$ we associate to the Wilson Dirac operator 
the hermitian 
Wilson Dirac operator, $H_W (m)=\gamma_{d+1} D_W (m) $.

All our lattices are assumed finite an therefore all our operators
are finite dimensional matrices. 
An eigenvalue of a matrix $A$
will be denoted by $\lambda (A)$; if the eigenvalues are labeled, the
label is attached to $\lambda$. When it makes sense, we may deal
with the maximal(minimal) eigenvalues of $A$, $\lambda_{\rm max(min)} (A)$.
We choose the following norm definition for matrices $A$: $\| A\|=
[\lambda_{\rm max} (A^\dagger A ) ]^{1\over 2}$. This is a standard choice,
induced by the vector norm $\| v \|^2 = \sum_{I} |v_I|^2$, where $I$ is
a generic component index [\horn]. 
The norm of a gauge covariant matrix is gauge 
invariant. 
\vskip .5cm
{\bf Formal Continuum Limit}
\vskip .5cm
The connection to the continuum is as follows: Assume to be given smooth 
functions\footnote{${}^{f_1}$}{In general the $A_\mu(x)$ aren't smooth functions, 
rather they make up a  
one form $\sum_\mu A_\mu (x) dx_\mu$ which is a smooth
connection on a possibly
nontrivial bundle with structure group $SU(n)$ over the four-torus.
}
$A_\mu (x)$ on the torus. Then,
$$
\eqalign{
U_\mu (x)&= \lim_{N\to\infty} \left [ e^{i{a\over N}
A_\mu(x)}e^{i{a\over N}A_\mu(x+{a\over N}\hat\mu)} 
e^{i{a\over N}
A_\mu(x+2{a\over N}\hat\mu)} \dots e^{i{a\over N}
A_\mu(x+(N-1){a\over N}\hat\mu)}\right ]
\cr &\equiv
P \exp [i\int_l dx_\mu A_\mu (x) ]~~~~~{\rm (
the~symbol~P~denotes~``path~ordering" )}.} 
$$
Consider a smooth function $\psi_c (x)$ with same index structure as
the corresponding object on the lattice. By looking at $x$'s coinciding
with a lattice point one gets a lattice vector $\psi (x=\vec n a)$, where
$\vec n \in Z^d$. 
The action of the $T_\mu$ produces another lattice vector $\psi^\prime$.
One can define a continuum operator, $T_{\mu c}$ such that the lattice
restriction of $T_{\mu c} \psi_c$ will be a function $\psi_c^\prime$
whose lattice restriction is $\psi^\prime$. The formula is\footnote{$
{}^{f_2}$}{This
generalizes an observation of van Baal [\vanbaal].}:
$$
T_{\mu c} = e^{aD_\mu},~~~~D_\mu =\partial_\mu + iA_\mu .$$
The simplicity of this expression can be viewed as a motivation
to introduce the $T_\mu$'s as central objects on the lattice in the
first place.  

The formula is easy to prove:
$$
\psi_c (x+a\hat\mu ) = e^{a\partial_\mu} \psi_c (x) .$$
for any vector $\psi_c$. On the other hand, for any operator ${\cal O}_c$
acting pointwise by ${\cal O}_c(x)$ we have:
$$
{\cal O}_c 
(x+b\hat\mu)= e^{b\partial_\mu } {\cal O}_c (x) e^{-b\partial_\mu } .$$
Inserting this expression (with $b=k{a\over N}$)
repeatedly into the definition of $U_\mu (x)$, implementing the shift
of the argument of $\psi_c (x)$ as above, 
and taking $N$ to infinity at the end, produces the desired
result using Trotter's formula [\reedsimon]. 

The $V_\mu$'s have associated continuum operators $V_{\mu c}$,
given by:
$$
V_{\mu c} = e^{- a\gamma_\mu D_\mu}~~~~~~~~{\rm no~sum~on~}\mu .$$
The Wilson Dirac operator is a lattice restriction of the continuum
operator
$$
a D_{W c} (m)=m+\sum_\mu (1-e^{-a\gamma_\mu D_\mu } ).$$
$D_{W c} (m)$ could be viewed as an 
approximation to $\gamma_\mu D_\mu$ in the continuum which
is good for eigenvalues small in absolute value but whose
spectrum is restricted to a bounded domain. Such operators
are frequently introduced when one regulates infinities in 
the continuum. 
The continuum Dirac operator $\sum_\mu \gamma_\mu D_\mu$ 
formally emerges as $a$ goes to zero,
and the mass is of order ${m\over a}$, where $m$ is a pure number. 
But, as an operator in
the continuum, $D_{W c} (m)$ is special: when it acts on $\psi_c$
to produce $\psi_c^\prime$, the values of $\psi_c^\prime$
at lattice points are solely determined by values of $\psi_c$
at lattice points. Therefore, there exists an exact relation to the
lattice operator $D_W (m)$.

There is no remnant of chiral symmetry (for even dimension $d$)
because $D_{W c} (m)$ isn't
just a function of $\sum_\mu \gamma_\mu D_\mu$; only in the small $a$
limit (strictly speaking, one would need to replace $m$ by $m_c a$ before
taking $a$ to zero) do we get an expression involving only
the chiral combination $\sum_\mu \gamma_\mu D_\mu$. 

It is important to appreciate that one does not need $D_{W c} (0)$ 
to anticommute
with $\gamma_{d+1}$ to have some amount of 
lattice chirality: any reasonable $D_c (m)$ that
is a function of only the combination 
$\sum_\mu \gamma_\mu D_\mu$ would do. For example,
if $aD_{W c} (m)$ were replaced by
$$
aD^\prime_{W c} (m) = m+1 -e^{\sum_\mu \gamma_\mu D_\mu} , $$
we would have enough symmetry because
$$
\gamma_{d+1} e^{-{1\over 2}\sum_\mu \gamma_\mu D_\mu } 
\left [ e^\mu - e^{-\mu + \sum_\mu \gamma_\mu D_\mu} \right ]
 e^{-{1\over 2}\sum_\mu \gamma_\mu D_\mu }\gamma_{d+1}=
-\left [ e^{- \mu} - e^{\mu + \sum_\mu \gamma_\mu D_\mu} \right ] .$$
Since $\det e^{-{1\over 2}\sum_\mu \gamma_\mu D_\mu }$ is unity  
${{\partial}\over {\partial \mu }}
\log \det \left [ e^\mu - e^{-\mu + 
\sum_\mu \gamma_\mu D_\mu} \right ]$ is odd in $\mu$ and this is
enough to eliminate additive quark mass renormalization.
However, the operator $e^{\sum_\mu \gamma_\mu D_\mu}$ cannot be restricted
to the lattice because when it acts on $\psi$ and produces $\psi^\prime$
it is not true that the values of $\psi^\prime$ at lattice points
depend only on values of $\psi$ at lattice points. 

One can try to 
``improve'' $D_{W c} (m)$ by looking at the difference $D^\prime_{W c} (m) 
-D_{W c}(m)$ to leading order in $a$ and replacing it by a function of
the $T_{\mu c}$ (again to leading order in $a$). Adding the new term
to $D_{W c} (m)$ produces an operator which can be restricted
to the lattice and is ``clover improved''; it
agrees with $D^\prime_{W c} (m)$ to leading and subleading order
in $a$. In fluctuating gauge field backgrounds
one changes the coefficient of the new term to a number determined numerically.

One can also maintain chiral symmetry on the lattice
exactly [\ovlap] [\wupp], using  
the overlap Dirac operator.

\vskip .5cm

{\bf Upper bound}
\vskip .5cm
Our first objective is to find a bound for the largest eigenvalue of $H_W^2$.
Clearly, $\lambda_{\max} (H_W^2 )=\|D_W (m) \|^2$. The triangle inequality
then gives:
$$\| D_W (m) \| \leq |m+d|+\sum_\mu \|V_\mu \| = |m+d| + d .$$
The lowest upper bound as a function of mass
is obtained at $m=-d$, which is a symmetry
point for $H_W^2 (m)$, because $D_W (-d)$ and $-D_W(-d)$ are
unitarily equivalent. This is a consequence
of the existence of a unitary hermitian operator
$S$ such that $SV_\mu S = -V_\mu$, implying $SD_W (m)S=
-D_W (-m -2d)$; $S$ is diagonal and the diagonal entries
are $1$ if the site $x$ has $\sum_\mu x_\mu$ even and $-1$
otherwise. $S$ exists because the hypercubic lattice we are
working on is bipartite.

For $m \geq -d$ the upper bound is attained iff there exits a 
vector $\psi$ which is a common eigenvector to all $d$ $V_\mu$
operators, with the eigenvalue $-1$ in each case. It is likely to
find such an eigenvector when $[T_\mu , T_\nu ] =0$ for all $\mu$
and $\nu$. These commutators vanish when all plaquette parallel
transporters are unity; this is so, in particular, in the free case.
\vskip .5cm
{\bf Lower bound}
\vskip .5cm
Let us introduce some shorthand notation:
$$
h_\mu = {1\over 2} (T_\mu + T_\mu^\dagger )=h_\mu^\dagger ,~~~~~~~
a_\mu = {1\over 2} \gamma_\mu (T_\mu^\dagger -  T_\mu )=-a_\mu^\dagger .$$
The unitarity of $V_\mu$ holds because of the identities
$$
h_\mu^2 - a_\mu^2 =1,~~~~~~~[h_\mu , a_\mu ] =0 .$$

Let $\lambda(m) =\lambda(H_W (m))$ be some eigenvalue of $H_W (m)$.
$\lambda (m)$ is differentiable because $H_W (m)$ depends smoothly on
$m$: ${{d\lambda}\over{dm}} = \sum_{x,i,\alpha,\beta }\psi^{i\ *}_\alpha (x)
\gamma_{5 \ \alpha,\beta} \psi^{i}_\beta (x)$, 
where $H_W (m) \psi = \lambda (m)
\psi$ and $\psi$ has unit norm. Since $\gamma_5^2 =1$ one has
$$
\left | {{d\lambda}\over {dm}} \right | \leq 1 .$$
The theoretical usefulness of expressions for ${{d\lambda}\over{dm}}$
has been recently emphasized by Kerler [\kerler].
This inequality restricts the slope of lines describing the flow
of eigenvalues of $H_W (m)$ as a function of $m$. We shall refer
to this inequality as the ``flow inequality''. It has an important
consequence that we shall prove below:
If we know that $0 < \lambda_{\rm min} (H_W^2 (m))$
for some $m$, we have 
$$
\left [ \lambda{\rm min} (H_W^2 (m^\prime ) ) \right ]^{1\over 2}
\geq \left [ \lambda{\rm min} (H_W^2 (m) ) \right ]^{1\over 2}-|m-m^\prime| .$$
Before describing the proof let us note 
that the result is useful only if 
$$
| m - m^\prime | 
< \left [ \lambda{\rm min} (H_W^2 (m) ) \right ]^{1\over 2} .$$
The main observation 
is that a lower bound on   
$\left [ \lambda{\rm min} (H_W^2 (m) ) \right ]^{1\over 2}$
at an arbitrary mass
point $m$ can be extended to a lower bound on
$\left [ \lambda{\rm min} (H_W^2 (m^\prime) ) \right ]^{1\over 2}$
in some mass
mass range around $m$.

The basic inequality can be best proven appealing to a sketch shown
in Figure 3:
The graphical meaning of the inequality is that $H_W^2 (m)$
has no eigenvalues in the area bounded by the right angle rhombus
in the figure when it is given that there are no eigenvalues
along its main diagonal $(A,B)$. Recognizing this, the
proof becomes trivial: if we did have an
eigenvalue anywhere inside the rhombus the flow inequality would have 
to be violated somewhere in order to avoid an eigenvalue flow crossing the
main diagonal. 

\figure{3}{\captionthree}{3.5}

We start with an explicit formula for $H_W^2 (m)$:
$$\eqalign{
H_W^2 (m) =& \left [ m+\sum_\mu (1-h_\mu ) \right ]^2 - \left [ \sum_\mu a_\mu
\right ]^2 -\sum_{\mu \neq \nu } [a_\mu , h_\nu ] =\cr
&m^2+2(m+1)\sum_\mu (1-h_\mu ) +\sum_{\mu\ne\nu} \left [
(1-h_\mu ) (1-h_\nu ) - a_\mu a_\nu -[a_\mu , h_\nu ] \right ] .}
$$
While in the continuum $D^\dagger (m) D (m)$ commutes with $\gamma_{d+1}$
the last term in $H_W^2 (m)$ does not. 
All terms are individually hermitian. Since $H_W(m)$
connected sites $x$, $x^\prime$ with\footnote{${}^{f_3}$}{
For two sites $x$ and $y$ we define $|x-y|=
\sqrt {\sum_\mu  (x_\mu - y_\mu )^2 }$} $|x-x^\prime |=0,1$
we could have expected $H_W^2 (m)$ to connect sites 
with $|x-x^\prime |=0,1,\sqrt{2},2$, but because of the relations
$h_\mu^2 - a_\mu^2 =1$ and $[h_\mu ,a_\mu ]=0$ 
sites with $|x-x^\prime |=2$ are still disconnected. Another
special property of $H_W^2 (m)$ is that the site diagonal piece
is proportional to the identity matrix. 

If $[T_\mu , T_\nu ]=0$ we have $ \sum_{\mu\ne\nu} a_\mu a_\nu =0$, 
$[h_\mu, a_\nu] =0$ and $[h_\mu , h_\nu]=0$. 
Then,
$$
H_W^2 (m) = m^2 + 2(m+1)\sum_\mu (1-h_\mu ) +\sum_{\mu\neq \nu } 
(1-h_\mu)(1-h_\nu) .$$
If we keep all $h_\mu$ fixed but one, say $h_\nu$, the dependence on the
latter is {\it linear}, so the extremal values are obtained
at $h_\nu =\pm 1$. The argument is applied again and again to a
decreasing number of remaining directions 
leading to the conclusion that in order to find the extrema of $H_W^2$,
viewed as a function of the quantities $h_\mu$ (more precisely, their
eigenvalues, since the $h_\mu$ can be simultaneously diagonalized
by assumption)
we only need to check the $2^d$ possibilities $h_\mu =\pm 1$. 
The upper bound comes out as above, and the lower bound
on $\left [ \lambda_{\rm min} (H_W^2 )\right ]^{1\over 2}$ has the
shape shown in the Figure 2.
We learn that at the points $m=0,-2, -4, \dots -2d$ the theory has massless
fermions; the multiplicities are given by 
${{d!}\over{(d-n)! n !}}$ where $m=-2n$, $n=0,1,2,\dots ,d$. Thus, for
$m=-2, -4,   \dots -2d+2$ we have several doublers, the number
of different species being given by the number of different $h_\mu$
configurations producing a zero at the respective special mass point.  

From now on we shall concentrate on the region $-2 < m < 0 $. 
This region is interesting when we want to deal with one Dirac fermion
and avoid doublers. The region close to $m=0$ is important for traditional
numerical QCD with Wilson fermions. The region close to $m=-1$
is important for applications of the overlap Dirac operator where one
would like $H_W^2 (m)$ to have a large gap around zero. 
When $[ T_\mu , T_\nu ] =0$, the highest lower bound is obtained at $m=-1$.
As long as all operators $[ T_\mu , T_\nu ]$ are small in norm we
expect the same to be true. We therefore focus on the point
$m=-1$ first, and later extend the bound to a range around $m=-1$
using the consequence of the flow inequality established earlier. 

In the general case where the matrices $T_\mu$ do not commute,
we have
$$
H_W^2 (-1) = 1+ \sum_{\mu \neq \nu} \left [ (1-h_\mu ) (1-h_\nu ) -
a_\mu a_\nu - [a_\mu ,h_\nu ]\right ] .$$
We now analyze each term in the bracket individually; 
we treat them separately because their spinorial index structures
are different. The first term is rewritten as: 
$$
\sum_{\mu \neq \nu} \left [ (1-h_\mu ) (1-h_\nu ) \right ] =
{1\over 4} \sum_{\mu \neq \nu} (1-T_\mu ) (1-T_\mu^\dagger ) (1-T_\nu )
(1-T_\nu^\dagger )= Q + X .
$$
Here, $Q$ is positive semidefinite,
$$
Q={1\over 8}\sum_{\mu \neq \nu} \left [ (1-T_\mu ) (1-T_\nu ) [ (1-T_\mu )
(1-T_\nu ) ]^\dagger + (1-T_\mu^\dagger ) (1-T_\nu ) [ (1-T_\mu^\dagger )
(1-T_\nu ) ]^\dagger\right ],$$
while $X$ depends only on $T_\mu$-commutators:
$$
X=-{1\over 8}\sum_{\mu \neq \nu} 
\left ( T_\mu [T_\mu^\dagger , T_\nu +T_\nu^\dagger ] + T_\mu^\dagger
[T_\mu , T_\nu +T_\nu^\dagger ] \right ) = 
-{1\over 8}\sum_\mu [T_\mu , [T_\mu^\dagger , 
\sum_\nu (T_\nu + T_\nu^\dagger ) ]] .$$
Proceeding, we find
$$
-\sum_{\mu \ne \nu } a_\mu a_\nu = -{1\over 8} \sum_{\mu\ne\nu} 
\gamma_\mu\gamma_\nu
[T_\mu - T_\mu^\dagger, T_\nu - T_\nu^\dagger ] = Y,$$
and
$$-\sum_{\mu\ne\nu} [a_\mu , h_\nu ] = {1\over 8}\sum_{\mu\neq\nu} 
\left [ (\gamma_\mu - \gamma_\nu ) ( [T_\mu , T_\nu ] + h.c.)
+(\gamma_\mu + \gamma_\nu ) ( [T_\mu , T_\nu^\dagger ] + h.c. )\right ] = Z.$$
The operators $Q,X,Y,Z$ are all hermitian. Moreover, each of the 
traces of $X^2 , Y^2 , Z^2$
are  linearly
related to the single plaquette Wilson action (see below) and decrease
when the latter increases and the continuum limit is approached.  

Consider now the commutators $[T_\mu , T_\nu ]$. Their norm is
determined by:
$$
[T_\mu , T_\nu ]^\dagger [T_\mu , T_\nu ]= (1-P_{\mu\nu})^\dagger
(1-P_{\mu\nu}),$$
where the unitary $P_{\mu\nu}$ are given by:
$$P_{\mu\nu} = T_\nu^\dagger T_\mu^\dagger T_\nu T_\mu .$$
The operators $P_{\mu\nu}$ are site diagonal,
with entries that are parallel transporters round plaquettes:
$$
(P_{\mu\nu}\psi)(x) = U_{\mu\nu} (x) \psi (x),~~~~~~~~~~U_{\mu\nu}(x)=
U_\nu^\dagger (x-\hat\nu ) U_\mu^\dagger (x -\hat\nu -\hat\mu ) 
U_\nu (x -\hat\nu - \hat\mu ) U_\mu (x-\hat\mu ) .$$
$U_{\mu\nu} (x)$ is associated with the elementary loop
starting at site $x$, going first in the negative $\nu$ direction,
then in the negative $\mu$ direction, and coming back round the plaquette. 

The main relation is:
$$\|[T_\mu , T_\nu ]\| = \|1 - P_{\mu\nu}\|.$$
Any pure gauge action with the right continuum limit will strongly
prefer configurations where all $U_{\mu\nu}(x)$ are close to unit matrix.
Therefore, it is not unreasonable to impose the constraint, for all
$\mu > \nu$, 
$$
\|[ T_\mu , T_\nu ]\| \leq \epsilon_{\mu \nu} .$$
Note that this is equivalent to
$$
\|1-U_{\mu\nu}(x)\| \leq \epsilon_{\mu\nu} $$
for every site $x$. 
It is easy to see that the same bound will hold
when we interchange in the commutator the $\mu, \nu$ indices,
and when we replace, independently, the $T_{\mu}$ and $T_{\nu}$ operators by
their hermitian conjugates. 

Using the triangle inequality and that $\|AB\| \leq \|A\|\|B\|$, 
we now obtain:
$$
\|X\| \leq \sum_{\mu > \nu } \epsilon_{\mu\nu},~~
\|Y\| \leq \sum_{\mu > \nu } \epsilon_{\mu\nu},~~
\|Z\| \leq \sqrt{2} \sum_{\mu > \nu } \epsilon_{\mu\nu} .
$$
The $\sqrt{2}$ factor comes in because $(\gamma_\mu \pm \gamma_\nu )^2 =2$
for $\mu\ne\nu$. We finally obtain:
$$
\lambda_{\rm min} (X) \geq -\sum_{\mu > \nu } \epsilon_{\mu\nu},~~
\lambda_{\rm min} (Y) \geq -\sum_{\mu > \nu } \epsilon_{\mu\nu},~~
\lambda_{\rm min} (Z) \geq -\sqrt{2} \sum_{\mu > \nu } \epsilon_{\mu\nu} .$$
By the variational principle and the positivity of $Q$ we arrive at:
$$
\lambda_{\rm min} (H^2 (-1) ) \geq 1 - (2 +\sqrt{2}) 
\sum_{\mu > \nu }\epsilon_{\mu\nu} .
$$
Our result is meaningful only when the number on the right hand side
in the above equation is non-negative. 

In the rotational invariant case one could set $\epsilon_{\mu\nu}=\eta$.
Then, for $d=4$, we obtain
$$
\sqrt {\lambda_{\rm min} (H^2 (-1) )} 
\geq \sqrt {1 - 6(2+\sqrt{2})\eta}\approx
\sqrt {1 - 20.5\eta} .$$

The general bound we obtained is:
$$
\left [ \lambda_{\rm min} (D_W^\dagger (m ) D_W (m))\right ]^{1\over 2}\geq
\left [ 1 - (2+\sqrt{2}) 
\sum_{\mu > \nu }\epsilon_{\mu\nu} \right ]^{1\over 2} -|1+m| .$$
This bound is useful only for 
$$
|1+m|\leq \left [ 1 - (2+\sqrt{2}) 
\sum_{\mu > \nu }\epsilon_{\mu\nu} \right ]^{1\over 2} .$$
This range is contained in the open segment
$-2 <  m < 0 $. The bound holds in both even and odd dimensions.
In the particular case of domain wall fermions, 
plaquettes parallel to the extra dimension
make no contribution since their $\epsilon_{\mu\nu}$ vanishes. 

\vskip .5cm
{\bf Comparison to other work}
\vskip .5cm
Related issues were studied in [\hernan] and in [\adams]. 
The authors of [\hernan] established the upper bound 
$$
\sqrt{\lambda_{\rm max} (H_W^2 (m) )} \leq 8 .$$
in four dimensions with the restriction $-2<m<0$. This is compatible, 
but less stringent than our upper bound, which becomes $m+8$ in this
mass range. 

In numerical investigations with pure gauge Wilson action, it was reported
in [\hernan] that, for $\beta = 6.0, 6.2, 6.4$ and $m=-1.0, -1.2, 
-1.4, -1.6$, for $SU(3)$, 
$\lambda_{\rm max} (H_W^2 (m) )$ stays around 41 and hardly changes. Our upper
bound for $m=-1.6$ is $6.4^2 = 40.96$ and increases for the lower $m$'s.
Thus, at the extremal mass value (assuming the value quoted
in [\hernan] was rounded), our bound is saturated to numerical accuracy. 
The claimed mass independence seems surprising, and not entirely
consistent with numerical results at other $\beta$ values and volume
sizes \footnote{${}^{f_4}$}{
The numerical work was carried out by the SCRI group, at the time 
consisting of Edwards, Heller and Narayanan.}.

In [\hernan] a bound on $\lambda_{\rm min} (H_W^2 (-1))$ is also established.
It is expressed in terms of a bound on the norm of the commutators, but
the precise definition of the norm used is not given. I shall assume it 
is the one adopted in this paper. 
A bound is quoted only for the $d=4$ case, for $m=-1$\footnote{${}^{f_5}$}
{According to David Adams [\adams], in unpublished work, 
the authors of [\hernan]
have extended their bound by using the triangle 
inequality to a range of $m$ values contained within
the segment $(-2, 0)$.} and for the rotational invariant case
$\epsilon_{\mu\nu}=\eta$. The bound derived in [\hernan] is:
$$
\lambda_{\rm min} (H_W^2 (-1) ) > 1-30\eta .$$
This bound is compatible with the result of this paper, 
$\lambda_{\rm min} (H_W^2 (-1)) \geq [1-6(2+\sqrt{2})\eta ]$, 
but weaker. To be sure that $H_W^2 (-1)$ has no zero eigenvalues
the bound in [\hernan] places a restriction on $\eta$ that is stronger
than ours by about one third. 
\vskip .5cm
{\bf Lessons}
\vskip .5cm
Let us first identify what about our results could have been expected
without any calculations. Clearly, we know that there will be some uniform
upper bounds on the spectrum just by virtue of compactifying momentum space
and because the $U_\mu (x)$'s are unitary. 
Moreover, once the free case is worked out and the spectral restrictions
of Figure 2 are derived, one knows that close to the continuum the structure
will be essentially similar even in the presence of nontrivial gauge fields.
The reasoning is as follows: We are dealing with operators that are
analytic in $T_\mu$ and  Figure 2 holds whenever all $T_\mu$'s commute.
All that enters in the bound derivations above is that the $T_\mu$'s
are unitary. Commuting unitary $T_\mu$'s can be smoothly deformed into
non-commuting ones and the changes in the spectrum must be smooth
too. Thus, if the commutators of the $T_\mu$'s are sufficiently
small there will be a region around $m=-1$ where
the spectrum of $H_W (m)$ will have a gap around zero. One can simply
think about the commuting case as a ``semiclassical'' approximation to
the non-commuting case. 

The operators $T_\mu$ connect only sites
one spacing apart in the $\mu$-direction. The gauge invariant norm
of the $T_\mu$ commutators cannot depend on anything else but the
norm of the elementary plaquettes. Forcing all unitary plaquette
operators close to identity produces a link configuration for which 
the $T_\mu$'s almost commute. The precise relation
between the $T_\mu$-commutators and the plaquettes is well known [\qek]
since the discovery of large $n$ reduction of lattice gauge theories [\ek]. 

So, all that really required some work was to turn the above into
a quantitative estimate. Because of the practical difficulties
associated with low eigenvalues of $H^2_W (m)$ it makes sense to
try to be as careful as possible in deriving the quantitative form
of the bounds. Still, it is known that the lower bounds in the $-2<m<0$
region are not directly useful in
backgrounds generated at coupling constants that are practical
in numerical QCD today. In spite of this, the exact bounds and their
derivation might provide helpful insights, in
particular in the context of implementations of the overlap Dirac 
operator. In this case one wishes to work with operators $H_W (m)$
with $-2 < m < 0$ but with as large a gap around zero as possible.
This would make the matrix $H_W (m)$ well conditioned and speed up
the calculations.

The most basic observation is that one can control the gap in $H_W (m)$
by controlling the plaquette variables alone\footnote{${}^{f_6}$}{This 
was exploited when the
parameters of the first dynamical simulation of the exactly massless
Schwinger model were chosen [\schwinger].}. This was understood
long ago [\boulder]; a natural guess would be that replacing the pure
Wilson gauge action by the so called ``positive plaquette'' model [\urspp]
(for gauge group $SU(2)$) will create a gap around zero. 
Numerical checks by Urs Heller in early 1998 have shown that this was
not the case [\urscheck]. In addition, one cannot just change the form
of a single plaquette action and get something useful in four
dimensions. The correlation length increases exponentially as the
plaquettes are forced to identity and physical realistic volumes
rapidly become totally impractical. 
A milder approach is therefore called for. There are a few possibilities.

First, one could use a more complicated 
action then a single plaquette one. The
idea is that a more complicated action might make the plaquettes close
to unity, but still keep the gauge fields sufficiently random so
that the correlation length does not exceed a few lattice spacings. 
The improvements observed in simulations using domain walls (which
can be viewed as a particular truncation of the overlap [\prddwf]) 
when one switches from Wilson to so called ``Iwasaki actions'' might
be a reflection of this mechanism [\columa]. 
A more systematic approach
would be to follow an approximate renormalization group trajectory [\stama],
where the correlation length is controlled, to regimes in the coupling
constant space where the single plaquettes are closer to unity.
A note of caution: the inclusion of the fermionic determinant
in the gauge measure may be important and a fix 
that works for quenched simulations may fail in the dynamical case [\columb].

Another
observation is that making only the plaquettes in some
directions close to unity would help. 
This only requires to increase one dimension of the lattice and there
is no exponential relation between this dimension and
the closeness of the time-like plaquettes to unity. In four
dimensions there
are other good reasons for working on asymmetric lattices [\morning], 
so this looks
like a cheap and attractive alternative worth exploring\footnote{${}^{f_7}$}
{K.-F. Liu has informed me already in September that his group is  
studying some physics
questions using the overlap on 
asymmetric lattices.}. In lower dimensions
than four the impact of going to asymmetric lattices would 
be even more pronounced. 

Yet another possibility is to filter out the
``roughness'' from the gauge
background seen by the fermions by replacing the link variables $U_\mu (x)$
by new link variables $U^{\rm APE}_\mu (x)$ which are functions
of the original link variables, transform the same way under gauge
transformations, but produce
plaquette variables closer to unity. Recent work has obtained such
``APE smeared''  $U^{\rm APE}_\mu (x)$ [\ape] 
with associated plaquettes extremely
close to unity [\steph]. 
Of course too much ``filtering'' may take 
the lattice theory at typical simulation parameters too far away from the
desired continuum limit of QCD\footnote{${}^{f_8}$}{
Too little filtering may provide no advantages:
for example, in a dynamical simulation of a two dimensional chiral
model [\chiraltd], modest filtering produced no gains.}. 
If this is true, one could also try a ``half smeared''
approach where only the links entering the ``Wilson mass term''
${1\over 2} \sum_\mu (T_\mu + T_\mu^\dagger )$ in $D_W (m)$ 
are APE smeared but the links entering the chiral 
part ${1\over 2} \sum_\mu \gamma_\mu (T_\mu - T_\mu^\dagger )$ are not, 
so the fermions are not insulated from the ultraviolet fluctuations
in the gauge field. Unfortunately this would spoil the 
relations $h_\mu^2 -a_\mu^2 =1$ 
and $[h_\mu , a _\mu ]=0$, so the consequences on the 
bounds are complicated. Also, the spinorial structure no longer
only involves the projectors ${1\over 2}(1\pm \gamma_\mu )$ which
causes some numerical overhead. 
Note however that with APE smearing the difference
between  $U^{\rm APE}_\mu (x)$ and  $U_\mu (x)$ goes to zero
when the original $T_\mu$ commutators go to zero. Therefore, some
bounds of similar structure to the bounds presented here would still
hold.

It is hoped that the analysis of this paper would prove helpful in guiding
our search for improvements in the gauge action and in the structure of
$D_W (m)$.

\vskip .5cm

{\bf Acknowledgments}
\vskip .5cm

My research at Rutgers is partially supported by the DOE under grant
\# DE-FG05-96ER40559. I wish to thank David Adams for exchanges regarding
the lower bounds. I am indebted to Urs Heller for providing
numerical information on the
spectrum of the Wilson Dirac Hamiltonian in gauge fields generated
both with the positive plaquette action and with the Wilson action.
I am grateful to Rajamani Narayanan for making
available to me numerical results about eigenvalue flows and about upper
bounds on $H_W (m)$.
\vskip .5cm
{\bf References}
\vskip .5cm
\item{\bf \SCRI .} R. G. Edwards, U. M. Heller, R. Narayanan, \Journal{\PRD}
{60}{034502}{1999}.
\item{\bf \horn .} R. A. Horn, C. R. Johnson, ``Matrix Analysis'', Cambridge
University Press, 1985.
\item{\bf \vanbaal .} M. Garcia Perez, A. Gonzalez-Arroyo, J.
Snippe, P. van Baal, \Journal{\NPB}{413}{535}{1994}.
\item{\bf \reedsimon .} M. Reed, B. Simon, ``Methods of Modern 
Mathematical Physics, I: Functional Analysis'', Academic Press, 1973.
\item{\bf \ovlap .} R. Narayanan, H. Neuberger,
\Journal{\PLB}{302}{62}{1993};
\Journal{\NPB}{412}{574}{1994};
\Journal{\NPB}{443}{305}{1995}; H. Neuberger,
\Journal{\PLB}{417}{141}{1998};
\Journal{\PLB}{427}{353}{1998}.
\item{\bf \wupp .} H. Neuberger, hep-lat/9910040.
\item{\bf \kerler .} W. Kerler, hep-lat/9909031.
\item{\bf \hernan .} P. Hern{\' a}ndez, K. Jansen, M. L{\" u}scher, 
\Journal{\NPB}{552}{363}{1999}.
\item{\bf \adams .} D. H. Adams, hep-lat/9907005; contribution to Chiral'99,
``Workshop on Chiral Gauge Theories'' Sept. 13- Sept. 18, 1999, Taipei.
\item{\bf \qek .} G. Bhanot, U. M. Heller, H. Neuberger, 
\Journal{\PLB}{113}{47}{1982}
\item{\bf \ek .} T. Eguchi, H. Kawai, 
\Journal{\PRL}{48}{1063}{1982}.
\item{\bf \schwinger .} R. Narayanan, H. Neuberger, P. Vranas,
\Journal{\PLB}{353}{507}{1995}.
\item{\bf \boulder .} H. Neuberger, \Journal{\LTP}{73}{697}{1999}.
\item{\bf \urspp .} J. Fingberg, U.M. Heller, V. Mitrjushkin,
\Journal{\NPB .}{435}{311}{1995}.
\item{\bf \urscheck .} U. M. Heller, unpublished, private communication. 
\item{\bf \prddwf .} H. Neuberger, \Journal{\PRD}{57}{5417}{1998}.
\item{\bf \columa .} L. Wu, hep-lat/9909117.
\item{\bf \stama .} QCD-TARO Collaboration: 
Ph. de Forcrand et. al., hep-lat/9910011.
\item{\bf \columb .} P. Vranas, hep-lat/9911002.
\item{\bf \morning .} C. Morningstar, M. Peardon, \Journal{\PRD}{56}{4043}{1997}.
\item{\bf \ape .} M. Albanese et. al., \Journal{\PLB}{192}{163}{1987}.
\item{\bf \steph .} M. Stephenson, C. DeTar, T. DeGrand, A. Hasenfratz, 
hep-lat/9910023.
\item{\bf \chiraltd .} Y. Kikukawa, R. Narayanan, H. Neuberger, 
\Journal{\PRD}{57}{1233}{1998}.

\vfill\eject\bye